# Enhanced User Interaction in Operating Systems through Machine Learning Language Models

Chenwei Zhang[1]*, Wenran Lu[1], Chunhe Ni[2], Hongbo Wang[3], Jiang Wu[4]

[1]*Electrical and Computer Engineering, University of llinois Urbana-Champaign, Urbana, IL, USA
[1]Electrical Engineering, University of Texas at Austin, Austin, TX, USA
[2] Computer Science, University of Texas at Dallas, Richardson, TX, USA
[3] Computer Science, University of Southern California, Los Angeles, CA, USA
[4]Computer Science, University of Southern California, Los Angeles, CA, USA
* Corresponding author: zchenwei66@gmail.com

## ABSTRACT

With the large language model showing human-like logical reasoning and understanding ability, whether agents based on the large language model can simulate the interaction behavior of real users, so as to build a reliable virtual recommendation A/B test scene to help the application of recommendation research is an urgent, important and economic value problem. The combination of interaction design and machine learning can provide a more efficient and personalized user experience for products and services. This personalized service can meet the specific needs of users and improve user satisfaction and loyalty. Second, the interactive system can understand the user's views and needs for the product by providing a good user interface and interactive experience, and then use machine learning algorithms to improve and optimize the product. This iterative optimization process can continuously improve the quality and performance of the product to meet the changing needs of users. At the same time, designers need to consider how these algorithms and tools can be combined with interactive systems to provide a good user experience. This paper explores the potential applications of large language models, machine learning and interaction design for user interaction in recommendation systems and operating systems. By integrating these technologies, more intelligent and personalized services can be provided to meet user needs and promote continuous improvement and optimization of products. This is of great value for both recommendation research and user experience applications.

## 1. INTRODUCTION

"In the era of ever-expanding machine learning models, exemplified by giants like GPT with billions of parameters, there arises a pressing concern – are these colossal models truly learning as intended, or are they simply memorizing correct answers without comprehending underlying concepts, thereby struggling to adapt to novel problem scenarios? In response to this pivotal challenge, the field of explainable machine learning has emerged as a crucial frontier. In this paradigm, machine learning models not only make predictions but also provide transparent explanations to users, subsequently engaging in a dialogue to address questions and concerns.

Interactive machine learning (IML), as an innovative approach, not only showcases the interaction process but also incorporates human judgment and preferences into the development of machine learning models. This marriage of machine and human intelligence not only enhances user trust and preference for machine learning models' behavior but also points towards a promising direction for integrating human factors into the field of machine learning. While this work places significant emphasis on the automated performance of Human-Machine Interaction (HMI), it unequivocally reaffirms the central role of the end user—the individual engaging with the technology—in shaping and advancing the development of machine learning systems. This comprehensive toolbox of methods underscores the preference and trustworthiness of IML over non-IML approaches, reinforcing the significance of user-centric design in technology development."

## 2. RELATED WORK

### 2.1 Machine learning (ML) interacts with the operating system

Machine learning (ML) models are increasingly making important decisions in several key areas, such as healthcare, finance, and law. However, the most advanced ML models, such as deep neural networks, become more complex and difficult to understand. This poses challenges for users of models in practical applications, as they need to understand why models make predictions and whether they can be trusted. As a result, users often turn to essentially explainable ML models because people can understand them more easily. However, black-box models are generally more flexible and accurate, which has encouraged the development of post-hoc explanations that explain the predictions of trained ML models. These interpretability techniques either fit a faithful model in the local area around the prediction, or examine internal model details, such as gradients, to interpret the prediction. However, recent research has shown that practitioners often struggle to use interpretability techniques because of the difficulty of figuring out which interpretations should be implemented, how to interpret them, and answering follow-up questions beyond the initial interpretation. Overall, understanding ML models through simple and intuitive interactions is a key bottleneck in many areas of machine learning applications.

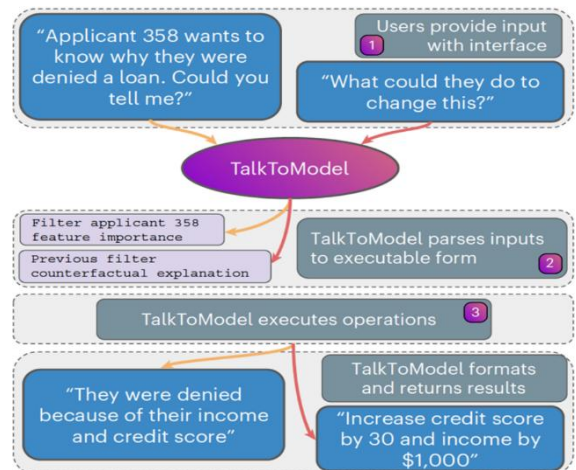

Figure 1.An overview of TalkToModel

In this paper, the authors address these challenges by introducing TalkToModel, a system that supports open natural language conversations for understanding ML models of any tabular dataset and classifier (Figure 1 provides an overview of TalkToModel). Users can talk to TalkToModel about why the forecast occurred, how the forecast will change if the data changes, and how to change the forecast, among many other conversation topics. In addition, they can perform these analyses on any group in the data, such as a single instance or a specific group of instances. For example, in a disease prediction task, a user could ask "How important is BMI for prediction?" Or "So how would lowering blood sugar levels by 10 percentage points change the likelihood of men over 20 developing the disease?". TalkToModel will respond by describing how BMI is the most important feature predicted and that lowering blood sugar will reduce the chance of diabetes by 20 percent, among other things. From here, users can further participate in the conversation by asking follow-up questions.

### 2.2 TalkToModel interacts with the user

TalkToModel is an interactive AI model based on natural language processing that enables dialogue and communication with users. Users can ask TalkToModel questions, ask for advice, request explanations, and so on, and the model responds in text and tries to understand and satisfy the user's needs. The way to interact with the user usually involves text input, but it can also be done through ways such as speech recognition.

1. User questions and requests: Users can submit various questions and requests to TalkToModel through text input. This can cover a variety of topics such as science, technology, culture, entertainment, health, education, and more. Users can ask factual questions, seek advice, request explanations of complex concepts, discuss topics, and more.

2. Model reply: TalkToModel will generate text reply based on the user's questions and requests. Responses usually contain relevant information, suggestions, explanations, or conversations with the user. The model tries to understand the user's needs and provide useful answers.

3. Interactivity: Interactions with users are usually continuous. Users can continue to ask questions, clarify questions, and delve into a topic, and the model responds to user input based on context, resulting in meaningful conversations.

"After a certain amount of user interaction, the user's interaction with the AI model is optimized through personalized recommendations. This leads to responses and suggestions that are better aligned with the user's interests and needs, ultimately enhancing the user experience, improving problem-solving efficiency, and fostering deeper and more meaningful interactions."

## 2.3  Machine learning language model

At present, many large or small domestic companies have begun to actively invest in and develop their own open source powerful language models, such as Ali Tongyi M6 large model and Tencent mixed Yuan are parameter capacity of more than 100 billion language models, and Baidu's word is also the first semantic large model terminal application. At present, there is no application that can directly compare Chat GPT in China, but the domestic language model can be better oriented to domestic users and surpass Chat GPT in terms of phrasing, quoting and Chinese language generation. The large language model is about to usher in rapid development in China, and the competition between companies will become increasingly fierce from the second half of this year.

Machine learning language models have evolved tremendously over time, with representative examples being large models like GPT-3, which have billions of parameters. Through deep learning technology, these models learn the structure and semantics of language from massive text data, enabling them to generate text, understand natural language, perform natural language processing tasks in a similar way to humans, and be applied in various fields, such as automatic translation, text generation, intelligent customer service, and so on.

The range of applications of machine learning language models is very broad, and they are changing the way many industries and fields are used. From a natural language generation perspective, they can be used to create automated content, enhance writing tools, provide intelligent search and recommendations, and even assist in the fields of education and healthcare. In addition, these models can also be used for sentiment analysis, text classification, information retrieval and other tasks, providing enterprises, research institutions and developers with more powerful natural language processing tools, promoting the development of intelligence and automation.

## 2.4  The importance of operating system user interaction

As the cornerstone of the digital age, the operating system provides a running and supporting platform for all computer software. As a kind of software, the large model itself also runs on top of the operating system. Universal Large model Explore more application scenarios through the operating system. The universal large model has powerful language understanding and processing capabilities, which can provide intelligent experience for users. And the emergence of large models brings new interaction patterns. Through natural language processing and deep learning, the operating system can better understand the needs and intentions of users, and in the future, each user will have a dedicated AI assistant that will run on top of the operating system. The access of large models brings unlimited innovation space for the application development of operating systems. Through the large model, developers can design more intelligent applications for the operating system, such as intelligent search, recommendation system, automated office assistant, etc., thus enriching the application ecology of the operating system.

Therefore, in the future, AI will be one of the basic capabilities of the operating system, and partners can directly call the AI capabilities provided by the operating system to release the value of the platform. The impact of large models on the operating system also includes the modularity and customization of the system, as well as data security and privacy protection. It can also provide customized solutions for different users and enterprises. Through the application of large models, the operating system can better meet the needs of different scenarios, and achieve a more flexible and modular system architecture. All in all, with the spread of large models and AI technologies, data security and privacy protection have become particularly important. Future operating systems will need to provide powerful AI capabilities while keeping user data safe.

## 2.5 Deep learning technology interacts with the operating system

(1) Enhanced user interface

Deep learning has a great enhancement effect on the user interaction of the operating system, one of the most important step is to enhance the user interface through convolutional neural networks, thus extending the utilization of the operating system.

In the process of user interface enhancement, in Conformer: Convolution enhanced Transformer, how to apply convolution in Transformer is introduced. In this paper, how to apply the attention mechanism in Transformer to convolutional neural network is introduced.

For an image, its shape is generally [H, W, Fin], and a violent way to do this is to flatten H and W into [H * W, Fin], and then run attention directly on this matrix to get:

$$O_h = Softmax(\frac{(XW_q)(XW_k)^T}{\sqrt{d_k^h}})(XW_v) \qquad (1)$$

Where X is the flattened matrix of H and W. The results of bull attention need to be pieced together:

$$MHA(X) = Concat[O_1,…,O_{Nh}]W^O \qquad (2)$$

However, the above calculation method completely ignores the position information, resulting in the same result if the position of each pixel on the image is mixed and then attention can be obtained. The formula is as follows:

$$MHA(\Pi(X)) = \Pi(MHA(X)) \qquad (3)$$

Among them, π is a method of arrangement.

And we know that the success of convolution in images is strongly related to its ability to capture structural information. So location information can't be discarded. Therefore, similar to the 1-dimensional relative position described in Transformer's relative position coding, we use 2-dimensional relative position coding here.

More specifically, when calculating the logits of attention between position (ix, iy) and position (jx, jy), the formula is as follows:

$$L_{i,j} = \frac{q_i^T}{\sqrt{d_k^h}}(k_j + r_{j_x - i_x}^W + r_{j_y - i_y}^H) \qquad (4)$$

Notice that a relative position code is defined here for the x and y dimensions respectively. Thus, the formula for calculating attention becomes:

$$O_h = Softmax(\frac{QK^T + S_H^{rel} + S_W^{rel}}{\sqrt{d_k^h}})V \qquad (5)$$

Where, SHrel[i, j] = qirjx-ixH, SWrel is the same.

Similar to the one-dimensional relative position, the embedding of the relative position only looks at the relative position difference and has nothing to do with the absolute position.

Thus, Conformer is a deep learning model that combines Transformer and convolutional neural networks, and it plays an important role in user interface enhancement of operating systems. By introducing a convolutional layer, Conformer is able to better handle images and graphical interfaces to provide a more intuitive and rich user interaction experience. The convolutional layer helps capture spatial features and patterns in images, allowing Conformer to more accurately recognize user gestures, ICONS, and other visual elements, improving the responsiveness and accuracy of the interface.

In addition, Conformer provides powerful processing capabilities for natural language input through its Transformer section. This allows the operating system to better understand the user's language instructions and text input, which provides more efficient text processing, speech recognition, and natural language generation. Because Conformer combines the strengths of Convolution and Transformer, it achieves superior performance in user interface enhancement and natural language processing. This comprehensive capability enables the operating system to better meet the needs of users, provide intelligent and highly personalized user experience, and thus improve the interaction between the operating system and users. In summary, the Conformer model brings more advantages to the operating system through its convolution enhanced Transformer structure, improving the interactivity and user experience of the user interface.

## 3. METHODOLOGY

Recommendation systems play an increasingly important role in today's web applications, whether it is online shopping, news apps, social media or movie websites, personalized content recommendations have become a key component of the user experience. With the explosive growth of information, it becomes more and more difficult for users to find interesting content in the massive information, which is the problem of information overload. To solve this problem, recommendation systems came into being to provide users with personalized recommendation content by analyzing their historical behavior and interests, making it easier for users to discover what they might like.

This experiment aims to explore how to improve the performance and user interaction experience of movie recommendation system by combining large language model and convolutional neural network (CNN). We will use the MovieLens dataset, which contains a large number of user ratings and feedback on movies. Our goal is to build a personalized movie recommendation model that will introduce large language models that can understand and encode text descriptions of movies and create rich semantic representations for users and movies. By combining these representations and convolutional neural networks, we hope to improve the accuracy and diversity of personalized recommendations, thereby enhancing the user's interactive experience.

### 3.1 Data preparation

Load the MovieLens dataset, including user ratings and movie information.

The dataset is divided into three files: user data users.dat, movie data movies.dat, and rating data ratings.dat. User data has fields such as user gender, age, occupation ID, and zip code. There are 6040 user ids, gender is divided into two categories, age is divided into numerical values, and occupational ids are 19 categories. Movie data has fields such as movie ID, movie name, and movie style. There are 3883 movie ids, corresponding to 3883 movie names (including the year), there are 18 categories of Chinese film styles, and a movie has one or more movie styles. The rating data has fields such as user ID, movie ID, rating, and timestamp. Represents the ratings of the 6040 users for multiple movies, which are divided into five stars. The main goal of the experiment was to fit the user's rating of the movie. The data format is as follows:

Table 1. timestamps-Margins and print area specifications.

|   | UserID | MovieID | Rating | timestamps |
|---|--------|---------|--------|------------|
| 0 | 1 | 1193 | 5 | 978300760 |
| 1 | 1 | 661  | 3 | 978302109 |
| 2 | 1 | 914  | 3 | 978301968 |
| 3 | 1 | 3408 | 4 | 978300275 |
| 4 | 1 | 2355 | 5 | 978824291 |

Table 2. Genres-Margins and print area specifications.

|   | MovieID | Title | Genres |
|---|---------|-------|--------|
| 0 | 1 | Toy Story (1995) | Animation\|Children's\|Comedy |
| 1 | 2 | Jumanji (1995) | Adventure\|Children's\|Fantasy |
| 2 | 3 | Grumpier Old Men (1995) | Comedy\|Romance |
| 3 | 4 | Waiting to Exhale (1995) | Comedy\|Drama |
| 4 | 5 | Father of the Bride Part II (1995) | Comedy |

Preprocessing :Gender field: need to convert 'F' and 'M' to 0 and 1; Age field: To be converted into seven consecutive digits 0 to 6. Genres field: A category field that is turned into numbers. First turn the categories in the Genres into a string to a dictionary of numbers, and then turn the Genres field for each movie into a list of numbers, because some movies are a combination of multiple Genres. Title field: Works much like a Genres field, first creating a dictionary of text to numbers, then turning the description in Title into a list of numbers. The year in Ttle also needs to be removed. Genres and Title fields need to be uniformly long so that they can be easily processed in neural networks. The blanks are filled in with numbers corresponding to '<PAD>. The knowledge points used are sorted out like

(1)pandas.filter(): filter pandas data; map(dict) maps data to a dictionary. match().group() Matches and adds a group

(2) Convert the movie genre into a list of equal numbers, the length of which is 18. Movie names do much the same, as follows

genres _ map = {val: [genres2int[row for row in val.split('|')] for il,val in enumerate(set(movies['Genres'])}for key in genres_map:

for cnt in range(max(genres2int.values())-len(genres map[key])):genres maplkey].insert(len(genres maplkeyl)+ cnt,genres2int['<PAD>']).

### 3.2 Data preprocessing

Preprocessing involves converting categorical fields like Gender and Age into numerical representations, with Gender converted to 0 and 1 and Age to consecutive digits from 0 to 6. Genres and Titles, being category fields, are transformed into dictionaries of numbers. For Genres, each movie's list of genres is converted into a list of corresponding numbers, accommodating movies with multiple genres. Titles undergo similar treatment, where the description is converted into a list of numbers, excluding the year. Both Genres and Titles are standardized in length to facilitate neural network processing, with '<PAD>' used to fill any blanks.

Embedding layers are employed for UserID and MovieID to mitigate sparsity issues and dimensionality explosion. Each field is converted into numbers, serving as indices for the embedding matrix. The embedding layers have dimensions of (N, 32) and (N, 16) respectively. For movies with multiple genres, the embedding indices form a matrix (n, 32), which is then summed into a vector (1, 32).

The textual information in movie titles is handled using text convolutional networks instead of recurrent neural networks, streamlining processing. Post-indexing, features from the embedding layer are fed into fully connected layers, yielding two feature vectors of (1, 200) representing user and movie features.

Convolutional network model:

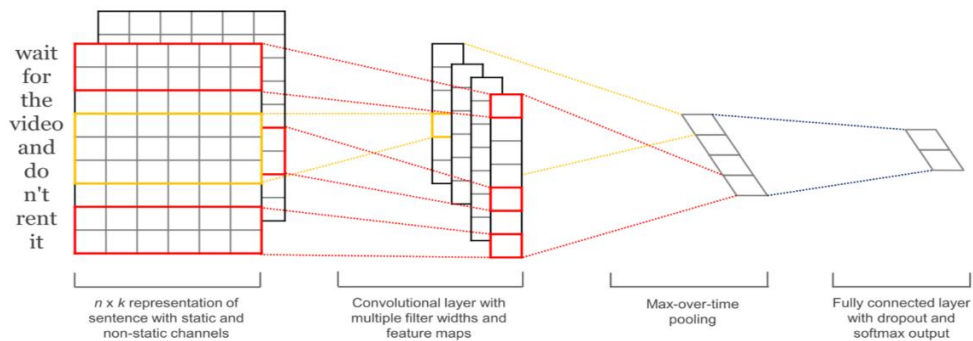

Figure 2. Convolutional neural network model

The first layer of the network is the word embedding layer, which is an embedding matrix composed of embedding vectors for each word. The next layer uses multiple convolution kernels of different sizes (window size) to do convolution on the embedded matrix, and window size refers to how many words each convolution covers. This is not quite the same as convolution for images, where convolution for images is usually 2x2, 3x3, 5x5, etc., whereas text convolution covers the entire embedding vector of words, so the size is (number of words, vector dimension), such as sliding 3, 4, or 5 words at a time. The third layer of the network uses max pooling to obtain a long vector, and finally regularizes it by using dropout to obtain the features of the movie Title.

### 3.3 Construct neural network

In this experiment, the core of the embedding matrix that defines a User is to map each user into a low-dimensional vector space so that these embeddings can be used in the model to represent the characteristics and preferences of the user. The core steps of this process include:

1. User ID code: Each user has a unique user ID, which is used to identify the user. In the experiment, the user ID will be used as part of the input data.

2. User embedding Matrix: Define a user embedding matrix, which is a matrix where each row corresponds to a user and each column corresponds to a different dimension of the user embedding vector. The number of rows in a matrix is usually equal to the total number of users, and the number of columns is equal to the dimensions embedded by the users.

3. Embedding layer: In the deep learning model, the embedding layer is used to map the user ID encoding to the corresponding user embedding vector. This map is learnable, and it updates the parameters in the embedded matrix through a backpropagation algorithm to minimize the model's loss function.

4. User embedding vector: Once the model is trained, each user is mapped to a user embedding vector that contains information about the user's characteristics and preferences. This user embedding vector will be used in the input layer of the model to represent the user's personalization information.

The core of the user embedding matrix is to capture the characteristics of the user by learning the embedding of the user ID, so that the model can understand and distinguish the differences between different users. This approach allows the model to personalize recommendations for each user, improving the accuracy of the recommendation system and user interaction experience.

### 3.4 Training network

```
%matplotlib inline
%config InlineBackend.figure_format = 'retina'
import matplotlib.pyplot as plt
import time
import datetime
losses = {'train':[], 'test':[]}
```

Its core principle is that the construction of neural network combines large-scale language model and convolutional neural network, makes full use of text information and user behavior data, and improves the performance of personalized movie recommendation system. This combination enables the model to better understand user interests and movie characteristics, achieve more accurate and diversified recommendations, while improving the user's interactive experience, solving the cold start problem in traditional collaborative filtering methods, and providing users with more attractive recommendation content.

### 3.5 TensorBoard Data result

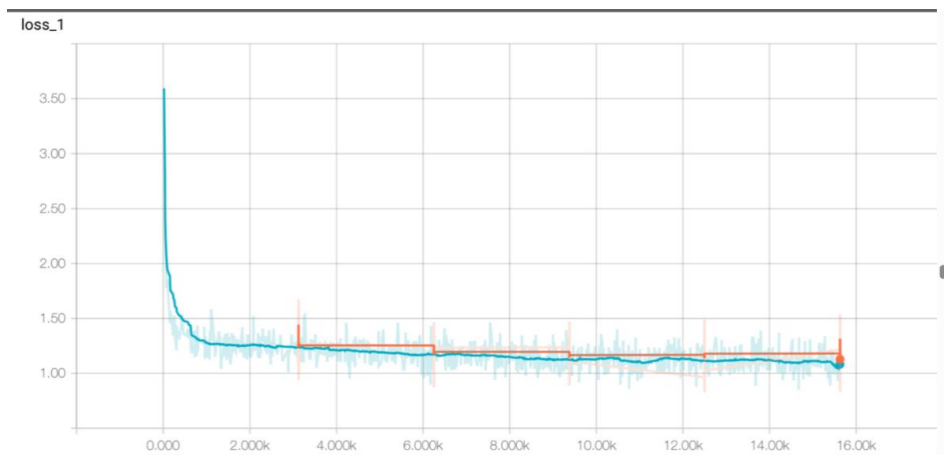

**Figure 3.** Model result diagram

After visualizing the model with TensorBoard, the next step is to analyze and interpret the model's performance and behavior. First, you can look at the model's loss curve and training metrics to evaluate the model's convergence and performance. Second, with visualization tools, it is possible to delve deeper into the internal structure of the model, looking at the activation and weight distribution of different layers to understand how the model handles the data. This helps identify potential problems and improve the structure of the model. The advantage is that by visualizing the results, the model can be better understood and adjusted, thereby improving its performance and interpretability and further optimizing the user experience.

In addition, visualization can also visualize the distribution and feature importance of input data, helping to find data biases and anomalies. This facilitates data preprocessing and model improvement. In addition, visualization helps to work with teams to share model results and insights, facilitating deeper discussion and decision making. Together, TensorBoard's visual results provide an opportunity to deeply understand and improve machine learning models, helping to improve their performance, interpretability, and collaboration efficiency.

## 4. CONCLUSION

This paper explores the integration of large language models, machine learning, and interaction design in user interaction and recommendation systems. By leveraging these technologies, more intelligent and personalized services can be delivered, enhancing user satisfaction and loyalty. The advancement of large language models enables machines to better comprehend and respond to user needs, opening new avenues for recommendation systems and operating systems. Visual tools like TensorBoard facilitate model analysis and refinement, further optimizing the user experience.

Future Perspective: Looking ahead, as large language models and machine learning continue to evolve, user interaction and recommendation systems will become increasingly intelligent and personalized. The operating system will serve as the foundation for tailored user services, while large models will offer developers ample opportunities for innovation, including intelligent search, recommendation systems, and automated office assistants. Additionally, data security and privacy protection will become paramount, with operating systems needing to deliver robust AI capabilities while safeguarding user data. In summary, large language models and machine learning techniques hold immense promise and challenges for user interaction and recommendation systems, poised to elevate user experiences, enhance recommendation accuracy, propel operating system development, and deliver more intelligent and personalized services to users.

## REFERENCES

[1] Vaswani, A., Shazeer, N., Parmar, N., Uszkoreit, J., Jones, L., Gomez, A. N., ... & Polosukhin, I. (2017). Attention is all you need. In Advances in neural information processing systems (pp. 30-38).
[2] Tianbo, Song, Hu Weijun, Cai Jiangfeng, Liu Weijia, Yuan Quan, and He Kun. "Bio-inspired Swarm Intelligence: a Flocking Project With Group Object Recognition." In 2023 3rd International Conference on Consumer Electronics and Computer Engineering (ICCECE), pp. 834-837. IEEE, 2023.DOI: 10.1109/mce.2022.3206678
[3] Liu, B., Zhao, X., Hu, H., Lin, Q., & Huang, J. (2023). Detection of Esophageal Cancer Lesions Based on CBAM Faster R-CNN. Journal of Theory and Practice of Engineering Science, 3(12), 36–42. https://doi.org/10.53469/jtpes.2023.03(12).06
[4] Liu, Bo, et al. "Integration and Performance Analysis of Artificial Intelligence and Computer Vision Based on Deep Learning Algorithms." arXiv preprint arXiv:2312.12872 (2023).
[5] He, X., Liao, L., Zhang, H., Nie, L., Hu, X., & Chua, T. S. (2017). Neural collaborative filtering. In Proceedings of the 26th international conference on world wide web (pp. 173-182).
[6] Zhang, L., Yao, L., Tay, Y., Chua, T. S., & Zhang, J. (2019). Deep learning based recommender system: A survey and new perspectives. ACM Computing Surveys (CSUR), 52(1), 1-38.
[7] Liu, Bo, et al. "Integration and Performance Analysis of Artificial Intelligence and Computer Vision Based on Deep Learning Algorithms." arXiv preprint arXiv:2312.12872 (2023).
[8] Yu, L., Liu, B., Lin, Q., Zhao, X., & Che, C. (2024). Semantic Similarity Matching for Patent Documents Using Ensemble BERT-related Model and Novel Text Processing Method. arXiv preprint arXiv:2401.06782.
[9] Jbene Mourad, et al. "Personalized PV system recommendation for enhanced solar energy harvesting using deep learning and collaborative filtering." Sustainable Energy Technologies and Assessments 60. (2023):
[10] Zafar Muhammad HamzaFalkenberg Langås Even,and Sanfilippo Filippo. "Empowering human-robot interaction using sEMG sensor: Hybrid deep learning model for accurate hand gesture recognition." Results in Engineering 20. (2023):